\documentclass[12pt, draftcls, letterpaper, onecolumn, journal]{IEEEtran}

\IEEEoverridecommandlockouts

\usepackage[cmex10]{amsmath}
\usepackage{amssymb}
\usepackage{subfigure}
\usepackage{graphicx,epsfig}
\usepackage{cite}
\usepackage{url}

\newtheorem{thry}{\textbf{Theorem}}

\begin{document}
\title{How Much Multiuser Diversity is Required for Energy Limited Multiuser Systems?}

\author{Youngwook~Ko,~\IEEEmembership{Member,~IEEE,}
        Sergiy~A.~Vorobyov,~\IEEEmembership{Senior Member,~IEEE,}
        and~Masoud~Ardakani,~\IEEEmembership{Member,~IEEE}
\thanks{The authors are with the Department
of Electrical and Computer Engineering, University of Alberta,
Edmonton, AB, T6G~2V4 Canada, e-mails:
(\{ko,~vorobyov,~ardakani\}@ece.ualberta.ca).}
\thanks{{\bf Corresponding author:} Sergiy A. Vorobyov, Dept. of \
Electrical and Computer Engineering, University of Alberta,
9107-116~St., Edmonton, Alberta, T6G 2V4, Canada; Phone: +1 (780)
492 9702, Fax: +1 (780) 492 1811.}
\thanks{This work is supported in part by the Natural
Science and Engineering Research Council (NSERC) of Canada and the
Alberta Ingenuity Foundation, Alberta, Canada.}
}

\maketitle

\vspace{-1.7cm}
\begin{abstract}
Multiuser diversity (MUDiv) is one of the central concepts in
multiuser (MU) systems. In particular, MUDiv allows for scheduling
among users in order to eliminate the negative effects of
unfavorable channel fading conditions of some users on the system
performance. Scheduling, however, consumes energy (e.g., for
making users' channel state information available to the
scheduler). This extra usage of energy, which could potentially be
used for data transmission, can be very wasteful, especially if
the number of users is large. In this paper, we answer the
question of how much MUDiv is required for energy limited MU
systems. Focusing on uplink MU wireless systems, we develop MU
scheduling algorithms which aim at maximizing the MUDiv gain.
Toward this end, we introduce a new realistic energy model which
accounts for scheduling energy and describes the distribution of
the total energy between scheduling and data transmission stages.
Using the fact that such energy distribution can be controlled by
varying the number of active users, we optimize this number by
either (i)~minimizing the overall system bit error rate (BER) for
a fixed total energy of all users in the system or (ii)~minimizing
the total energy of all users for fixed BER requirements. We find
that for a fixed number of available users, the achievable MUDiv
gain can be improved by activating only a subset of users. Using
asymptotic analysis and numerical simulations, we show that our
approach benefits from MUDiv gains higher than that achievable by
generic greedy access algorithm, which is the optimal scheduling
method for energy unlimited systems.
\end{abstract}



\section{Introduction}
\label{sec:intro}

In wireless systems, unfavorable channel conditions remain the
main hinderance to achieving desirable system throughput or bit
error rate (BER). To overcome this problem in multiuser (MU)
wireless systems, resource scheduling strategies, which use
channel fading conditions as an opportunistic resource, have been
proposed \cite{knopp,vits_ti02}. Using the so-called opportunistic
transmission, advanced scheduling strategies along with
closed-loop designs \cite{cap_fad_ch_w_csi,yogo06} have been
developed in the literature (see
\cite{vits_ti02,ottersten_com07,kote_tw08} and references
therein). The gain obtained by such opportunistic transmission
methods is known as multiuser diversity (MUDiv) gain.

For multi-point--to--point single--input single--output (SISO)
wireless systems, the MUDiv gain was first studied in
\cite{knopp}. The information theoretic results have shown that
based on the optimal transmit power control, the overall system
throughput can be maximized by allowing only the `best' user in a
system to transmit at each time slot. For downlink MU systems, the
MUDiv has been recognized as an effective method of improving the
system performance measures such as spectral efficiency and
quality of service over multipath fading channels
\cite{vits_ti02}. As a result, MUDiv approaches have been adopted
in commercial systems, e.g., systems based on orthogonal frequency
division multiple access (OFDMA) \cite{flash-ofdm}.

Toward improving the MUDiv gain, various system performance
measures and their tradeoffs have been considered, as well as
various algorithms have been developed
\cite{dimic_ts05}--\cite{hammarwall_ts08:ottersten}. In
\cite{dimic_ts05}, the problem of multiuser downlink beamforming
based on the MUDiv has been studied. In \cite{yang_jvt07}, the sum
capacity caused by the MUDiv gain has been investigated with
respect to two important MU system performance measures such as
fairness and scheduling complexity. In \cite{chaporkar07}, the
delay--energy tradeoff in MUDiv systems has been analyzed. It has
been shown that the energy required for guaranteeing an acceptable
rate per user decreases at the cost of a longer delay. In
\cite{geal_icc04,liang_spawc08:gershman}, low complexity
scheduling strategies based on low rate channel feedback from
users to the base station (BS) have been developed. In
\cite{hammarwall_ts08:ottersten}, it has been argued that if the
limited feedback is used, then the use of instantaneous channel
norm feedback provides additional spatial channel information so
that the MUDiv gain can be exploited efficiently in time,
frequency, and space. However, in the existing literature, the
MUDiv has been investigated for the case of fixed transmit
resources, e.g., fixed transmit power and \emph{fixed number of
active users}.

Although it has never been discussed before, it is important to
note that the MUDiv gain relies on the total energy available at
the users and, therefore, depends on this energy, especially for
the energy limited systems. Specifically, for scheduling purposes
all users must share their own channel state information (CSI)
with the BS per each data transmission. Then, a portion of the
energy available at each user must be used primarily for
scheduling, while only the remaining energy can be used for actual
data transmission. Therefore, the important question is how to
distribute the limited total energy available at the users between
scheduling and actual transmission stages? If all users are active
(available for scheduling) at all times, the waste of the energy
used primarily for scheduling may be very significant. The latter
will reduce the system performance. On the other hand, if only a
small number of users is kept active per each data transmission,
the corresponding MUDiv can be insufficient that also leads to
system performance degradation. Therefore, the aforementioned
question can be reformulated as the following signal processing
question: how much MUDiv is required for MU systems?

In this paper\footnote{Some preliminary results of this work have
been published in \cite{ko_icc09}.}, we develop methods which aim
at maximizing the MUDiv gain in MU systems by exploiting a
realistic energy model. Unlike existing schemes, we consider also
the energy spent by users to make their CSIs available to the BS.
By bringing this \emph{inherent} energy usage into the picture, we
find that it is better to choose (schedule for data transmission)
the `best' user from a subset of users (referred to as the set of
active users) rather than among the entire set of users. The
intuition is that, if a small subset of users is required to send
their CSIs to the BS, more energy can be saved for actual data
transmission and better overall system performance can be
achieved. This is especially true for the energy limited systems.
Thus, there is an inevitable tradeoff between the MUDiv and the
energy saved for actual data transmission. Using this tradeoff, we
aim at finding the optimal size of the set of active users so that
either the total system bit error rate (BER) or the total energy
of the users is minimized under practical system constraints.
Using asymptotic analysis, we also study how much MUDiv can be
achievable in various special cases of interest.

The paper is organized as follows. The system model is introduced
and the problem is described formally in Section~II. In
Section~III, the systems performance measures such as BER, upper
bound on BER, and approximate BER are derived. Section~IV contains
the answer to the main question of the paper, that is, how much
MUDiv is required for MU systems, while Section~V provides some
further analytical analysis. Extension to the case of multiple
antenna MU systems is given in Section~VI. Section~VII presents
numerical results and is followed by conclusions in Section~VIII.

\section{System model and problem description}
\label{sec:ch}

\subsection{System model}
Let $\bar{K}$ mobile users communicate with the BS. It is assumed
for simplicity that each user as well as the BS is equipped with a
single antenna. Thus, we consider an MU SISO system. This
assumption, however, will be generalized to the case of multiple
antenna MU systems in Section~\ref{sec:ext_mu}, and it will be
shown that such generalization is straightforward.

Suppose that the wireless channel between user~$k$ and the BS is
flat fading. The received signal at the BS from user~$k$ can be
then represented as
\begin{equation}
x_k = h_k s_k + v_k, \quad k=1,\cdots,\bar{K} \label{eq:io_ht}
\end{equation}
where the information--bearing symbol $s_k$ is a Gray--coded
quadrature amplitude modulated (QAM) symbol\footnote{Note that the
approach can be easily extended to other modulations.} from a
fixed constellation of size $M$, $v_k$ is the complex--valued
zero--mean additive white Gaussian noise (AWGN) with unit
variance, i.e., $v_k\sim{\cal CN}(0,1)$, and $h_k\sim{\cal
CN}(0,\sigma_k^2)$ is the channel gain between user $k$ and the
BS. We assume that $h_k$, $\forall k$ are independent and known
perfectly at the BS.

One of the main concerns for scheduling in the heterogeneous MU
environments is fairness among users. Among various fairness
notions such as, for example, average throughput per user
\cite{vits_ti02}, variance of short--term throughput per user
\cite{kote_tw08}, users' channel accessing period
\cite{elliot_ccece02}, our concern, in this paper, is fairness in
terms of the equal user's probability of accessing the channel.
According to this fairness notion, the scheduling is called fair
if the channel accessing probabilities are equal for all users in
the MU system. To satisfy such fairness conditions, we use an
opportunistic scheduling (OS) scheme proposed in \cite{knopp}.
This scheme incorporates an average power control which is
instrumental for our further considerations of the energy
distribution between scheduling and transmission stages. According
to this scheme, a ratio of the actual signal-to-noise ratio (SNR)
to its own average is used for both scheduling and data
transmission. The aforementioned scheduling scheme (hereafter
referred to as greedy access (GA) scheme) gives equal chance to
all users for accessing the channel. Thus, we employ it in this
work.

Let user~$k$ employ the average power control of \cite{knopp}
assuming that the variance of $h_k$, i.e., ${\sigma}_k^2$, is
known to him. Thus, the power is allocated to symbol $s_k$ in
(\ref{eq:io_ht}) so to obtain a desired average receive power at
the receiver which must be the same for all users. Then, denoting
the desired average receive power for unit transmit power by
$\omega$, the corresponding transmit power at user~$k$ can be
written as
\begin{equation}
\label{eq:txpw_k} \lambda_{k}=\frac{\omega}{ {\sigma}_k^2} \lambda
\end{equation}
where $\lambda$ is the transmit power before employing the average
power control and $\omega \lambda$ is the desired average receive
power that is equalized for all users via the average power
control $\omega/{\sigma}_k^2$. Therefore, using (\ref{eq:txpw_k})
and instantaneous channel gain $|h_k|^2,$ the instantaneous
receive SNR at the BS from user $k$ can be written as
\begin{equation}
\label{eq:rho_k0} \rho_k \triangleq |h_k|^2 \lambda_{k} .
\end{equation}
Since the variance of the AWGN in (\ref{eq:io_ht}) is unit,
\eqref{eq:rho_k0} can be equivalently written as
\begin{equation}
\label{eq:rho_k} \rho_k = |\tilde{h}_{k}|^2 \omega \lambda
\end{equation}
where $\tilde{h}_k\sim{\cal CN}(0,1)$. Therefore, $\forall k,$ the
distribution of $\rho_k$ is the same.

Using (\ref{eq:rho_k}) as a scheduling metric, we consider the GA
scheme, where at a given time slot, the BS chooses only one out of
multiple users for transmission. The user selection criterion is
based on finding the user with the most favorable channel gain
versus its own average. That is, user $k^*$ is scheduled for data
transmission if
\begin{equation}
k^*= \arg\max\limits_k \rho_k .
\end{equation}

In practical MU environments, the system resources such as the
number of users in the system and the total energy available at
each user are usually limited. Under such system limitations, one
interesting question is how the existing limits on the total
available energy of all users should change the requirements on
the MUDiv of the system. Indeed, one of the well known access
schemes, i.e., the random access (RA) scheme (see
\cite{tse_fundamental}), suggests to select users for transmission
randomly one at a time. This scheme provides no MUDiv, i.e.,
$K=1$, and, therefore, requires no extra energy spending for extra
communications between the users and the BS at the scheduling
stage. On the other hand, the GA scheme improves the performance
of MU systems due to its ability to select the `best' user for
transmission from the entire set of available users of size
$\bar{K}$ \cite{knopp,vits_ti02}. The MUDiv of the GA scheme is
then $K = \bar{K}$. Unfortunately, in this case, the BS has to
know the CSIs of all users in the system in the scheduling stage
which requires additional energy spending. Therefore, if the total
energy of all users in the system is limited, the use of the GA
scheme may be very wasteful in terms of the energy spent at the
scheduling stage. It reduces the energy available for actual data
transmission that can lead to the system performance degradation.
Therefore, the main query of this work is how much MUDiv $K$ is
required to improve the MU system performance? In other words, how
many users should transmit their pilot symbols so to make their
CSIs available at the BS. Based on these CSIs at the BS, one of
the users is selected to access the channel.

The aforementioned query can be solved by finding an optimal
energy distribution between scheduling and data transmission,
i.e., by selecting the cardinality $K = |{\cal A}| \in [1, \,
\bar{K}]$ of a subset of active users ${\cal A}$ which participate
in the scheduling. Here $|\cdot|$ denotes the cardinality of a set
and the elements (users) of ${\cal A}$ are selected randomly in
the beginning of every time slot according to a uniform
distribution. Such random selection in each time slot is
considered in order to achieve fairness among users in terms of
equal channel accessing probability.

Toward this end, let us first write the energies used for
scheduling and data transmission as functions of $K$. Taking into
account the scheduling stage, the energy consumed by user~$k$ at
each time slot for both scheduling and data transmission can be
defined as
\begin{equation}
E_{T,k} \triangleq E_{s,k} + {\bf 1}(k=k^*) \hspace{0.1cm}
E_{d,k}, \quad \forall k \label{eq:usr_ergslot}
\end{equation}
where $E_{s,k}$ denotes the energy spent for scheduling, $E_{d,k}
= T_s \lambda_{k}$ is the energy spent for data transmission,
$T_s$ stands for the symbol duration, and ${\bf 1}(k=k^*)$ is the
indicator function which is equal to 1 if $k=k^*$ and 0 otherwise.\footnote{Note that without loss of generality,
$E_{s,k}=E_{s,j}$, $\forall k, j$ is assumed. It corresponds, for
example, to the practical situation when the codeword length of
the transmitted symbol is long, while the number of pilot bits is
relatively small.}
%
%
Then, the total energy of all users over the time interval during
which the average channel gains remain constant can be found as
the sum of $E_{T,k},$ $\forall k$ over many time slots $N$
covering the whole interval. Since all users have equal chance of
accessing the channel, at a given time slot, any user has access
to the channel with probability $1/\bar{K}.$ Then, it can be found
that during $N$ time slots, the energy used by each user for
individual data transmission is $\lambda_k T_s \cdot N/{\bar K}$
where $\lambda_k$ is the transmit power at user $k$ which is equal
to the $k$th user SNR under the assumption of the unit variance of
the AWGN in \eqref{eq:io_ht}. Therefore, asymptotically for large
$N$, we can write that for $K\leq\bar{K}$, the total energy is
\begin{align}
E_T^{K-GA} 
\triangleq N \left(\lambda_{1}T_s + \cdots + \lambda_{\bar{K}}T_s
\right)/\bar{K} + K E_f = E_d + K E_f \label{eq:sys_erg}
\end{align}
where the superscript $(\cdot)^{K-GA}$ stands for the GA among $K$
active users, $E_f \triangleq N E_{s,k}$ denotes the energy
consumed by each user for scheduling over $N$ time slots, $K E_f$
denotes the total energy consumed by all users for scheduling, and
$E_d$ stands for the energy used by all users for actual data
transmission.

Although \eqref{eq:sys_erg} is an asymptotic result, it is
applicable to practical setups. Consider the random variable $X$
corresponding to the actual number of time slots that a user is
accessing the channel over $N$ time slots. Then $X$ has a binomial
distribution with average $m_X=N/\bar{K}$ and standard deviation
$\sigma_X = \sqrt{N /\bar{K} (1-1/\bar{K})} \approx \sqrt{N /
\bar{K}}$ for large $\bar{K}$. Therefore, for $\sigma_X <
{m_X}/{10}$, we need, $N/\bar{K}>100$ which is the realistic case
in practical setups.

\subsection{Problem description}
\label{sec:prb_desc} Two different objectives can be considered
for selecting $K$: (i) minimization of the system BER and (ii)
minimization of the total energy consumed by all users in the
system. Although the users are not connected to the same energy
source, given the finite energies at individual users, the sum of
individual user energies also determines the total energy consumed
by all users. It is worth stressing here that for system
performance analysis in MU systems, the total energy consumed by
all users is more important than individual user energies because
the MUDiv gain depends on the number of users participating in
scheduling, and the energy which determines the MUDiv gain is the
total energy consumed by all users, rather than the individual
user energies.
In addition, assume that for given channel statistics of all
users, the energy consumption by each user over given time slot(s)
is fixed on average. Then, the individual user energies are also
fixed fractions of the total energy of all users on average (see
\cite{knopp}, \cite{vits_ti02}, \cite{kote_tw08}, and references
therein for similar observations for power or data rate). Since
the individual user energies are fixed fractions of the total
energy, by minimizing the total energy, the individual user
energies are also minimized.
%

\subsubsection{System BER minimization} In
this case, we aim at minimizing the system BER under a constraint
on $E_T^{K-GA}$. Therefore, $E_T^{K-GA}$ is a constant independent
of $K$, and it is straightforward to see that $E_d$ in
\eqref{eq:sys_erg} as well as $K E_f$ depend on $K$ since the
energy distribution between $E_d$ and $K E_f$ must be optimized by
selecting such $K$ that minimizes the system BER. Thus, for a
given total energy consumed by all users, we first express the
tradeoff between $E_d (K)$ and $K E_f$ as a function $K$. Let us
define the ratio $\alpha \triangleq E_d^{GA} /E_f$ where
$E_d^{GA}$ denotes the energy for data transmission consumed in
the generic GA scheme that holds $K = \bar{K}$ during all time
slots. Here, the superscript $(\cdot)^{GA}$ stands for the GA
scheme. Then, representing $E_f$ in terms of $E_d^{GA},$
$E_T^{K-GA}$ can be expressed as
\begin{equation}
\label{eq:sys_usrerg1} E_T^{K-GA} = K \alpha^{-1} E_d^{GA} + E_d .
\end{equation}
Due to the fact that in the generic GA scheme $K = {\bar K}$ in
all time slots, the total energy consumed by all users is a
constant (denoted by $E_T^{GA}$). Constraining
(\ref{eq:sys_usrerg1}) to be equal to $E_T^{GA}$, we obtain under
such energy constraint
\begin{equation}
\label{eq:sys_usrerg} \underbrace{ \underbrace{ K \alpha^{-1}
E_d^{GA} }_{= K E_f} + E_d }_{ =E_T^{K-GA} } = \underbrace{
\underbrace{ \bar{K}\alpha^{-1} E_d^{GA} }_{= \bar{K} E_f} +
E_d^{GA} }_{=E_T^{GA}}
\end{equation}
where the two terms on the right hand side represent the total
energy $E_T^{GA}$ consumed in the generic GA scheme. It can be
seen from (\ref{eq:sys_usrerg}) that if $K$ is selected such that
$K \leq \bar{K}$, then more energy remains after scheduling, i.e.,
$E_d (K) = E_T^{GA} - K E_f$. This extra energy can be assigned
for actual data transmission, and $E_d(K)$ can be expressed in
terms of $K$ as
\begin{equation}
\label{eq:sys_txerg} E_d(K) = \left( (\bar{K}-K)\alpha^{-1} + 1
\right) E_d^{GA} .
\end{equation}
Therefore, $E_d(K)$ benefits from the energy gain of $(\bar{K} -
K) \alpha^{-1} + 1$ if $K < \bar{K}$. On the other hand, assigning
more energy for scheduling $K E_f$ increases the MUDiv gain.
Therefore, there exists a tradeoff between $E_d(K)$ and $K E_f$,
and the question now is where to spend the available energy in
order to minimize the total system BER. One of the possibilities
is to find the optimal value of $K\leq\bar{K}$ which minimizes the
total system BER, while satisfying the constraint on the limited
total energy of all users.

\subsubsection{Minimization of the total energy consumed by all
users in the system} In this case, we aim at minimizing the total
energy consumed by all users under the constraint that the system
BER remains below a pre-determined threshold. In order to satisfy
the BER requirement, $E_d$ in \eqref{eq:sys_erg} must remain
constant (i.e., $E_d = E_d^{GA}$) for any number of active users
$K$. Therefore, the total energy $E_T^{K-GA}$ can be minimized by
selecting such $K$ which minimizes $K E_f.$ Therefore, in this
case, the total energy $E_T^{K-GA}$ is also a function of $K$.
More precisely, since $E_d=E_d^{GA}$ for all $K$, then, by
selecting $K$, $E_T^{K-GA}$ can benefit from saving the energy at
the scheduling stage. Therefore, $E_T^{K-GA}$ can be expressed
versus $E_d^{GA}$ as
\begin{equation}
\label{eq:sys_usrerg_K} E_T^{K-GA} = K E_f + E_d^{GA} = (K
\alpha^{-1} + 1) E_d^{GA} .
\end{equation}
Since $E_T^{GA} = (\bar{K} \alpha^{-1} +1) E_d^{GA}$ and
$E_T^{GA}$ is a constant in the considered energy
minimization-based problem, $E_d^{GA}$ in (\ref{eq:sys_usrerg_K})
can be expressed via $E_T^{GA}$ as $E_d^{GA}=(\bar{K} \alpha^{-1}
+ 1)^{-1} E_T^{GA}$. Using this relationship and
(\ref{eq:sys_usrerg_K}), $E_T^{K-GA}$ can be further expressed
versus $E_T^{GA}$ as
\begin{equation}
\label{eq:sys_usrerg_K2} E_T^{K-GA} = \frac{K \alpha^{-1} +
1}{\bar{K} \alpha^{-1} + 1} E_T^{GA} .
\end{equation}
It can be seen from (\ref{eq:sys_usrerg_K2}) that the energy
saving gain is $(K \alpha^{-1} + 1)/(\bar{K} \alpha^{-1}+1)$ if $K
< \bar{K}$. Therefore, the smallest possible subset of available
users which satisfies the system target BER requirements is
optimal in terms of providing the minimum $E_T^{K-GA}$.

In order to express the aforementioned problems of selecting
optimum number of active users formally, we first need to find an
expression for the system BER as a function of $K$.

\section{System performance measures}
\label{sec:sys_prf} In this section, we derive expressions for the
exact, upper bound (UB), and approximate BERs. The exact BER
expression provides the highest accuracy for choosing $K$.
However, it may require intense computations, which may not be
practical in real-time. Therefore, a simple UB expression for BER
is derived. The use of the UB BER expression instead of the exact
BER in our problem will guarantee that the system BER requirements
will be satisfied, but the resulting $K$ may be sub-optimal.
Therefore, approximate BER expressions, which require the minimum
computations, are also derived.

\subsection{Exact BER expression}
\label{sec:exPb_ga} The exact BER of the M-ary modulation over the
AWGN channel can be written as \cite{chha03}
\begin{equation}
{\rm
Pr}_b(M,\rho)=\sum_{i=1}^{\Theta_M}C_{M,i}Q(\sqrt{c_{M,i}\rho})
\label{eq:ber_awgn}
\end{equation}
where $Q(x)\triangleq\pi^{-1}\int_{0}^{\pi/2}{\rm e}^{-x^2 /
2\sin^2\theta} \mathrm{d}\theta$ is the error function. For a
Gray-coded square M-ary quadrature amplitude modulation (QAM), the
constants $\Theta_M$, $C_{M,i}$, and $c_{M,i}$ can be found in
\cite{chha03}.\footnote{Note that the BER of a Gray-coded coherent
M-ary phase-shift keying (PSK) modulation in AWGN channel can also
be expressed using (\ref{eq:ber_awgn}). However, for brevity, only
M-ary QAM modulation is considered here.}

For a given $\omega$, the average BER is given by
\begin{equation}
{\rm Pr}_b=\int_0^\infty {\rm Pr}_b(M,y) p_{\rho}(y) {\rm d}y
\label{eq:pb_rho}
\end{equation}
where $p_{\rho}(\cdot)$ is the probability density function (pdf)
of $\rho$.

Let $\rho^{K-GA}$ denotes $\rho$ for $K$-GA scheme, i.e.,
$\rho^{K-GA} \triangleq \rho_{k^*}$. Considering the average power
control, $\rho_k$, $\forall k$ are independent and identically
distributed (i.i.d.) random variables. Using this fact and
applying higher order statistics, the pdf $p_{\rho^{K-GA}}(y)$ can
be found, for a given $K$, as
\begin{equation}
p_{\rho^{K-GA}}(y)= K \frac{{\rm
e}^{-y/\Omega}}{\Omega}\left(\gamma\left(1,\frac{y}{\Omega}
\right)\right)^{K-1} \label{eq:pdf_rho_g}
\end{equation}
where $\Omega\triangleq\omega \lambda$, and $\gamma(1,x)
\triangleq(1-{\rm e}^{-x})$. Using (\ref{eq:pdf_rho_g}), the
average BER can be written as \cite{simon_alouini_00}
\begin{equation}
{\rm Pr}_{b,e}^{K-GA} (K) = \sum_{i=1}^{\Theta_M} C_{M,i} \pi^{-1}
K \int_0^{\frac{\pi}{2}} a \int_0^{\infty} {\rm e}^{-t}
\left(\gamma(1,at)\right)^{K-1} {\rm d}t {\rm d} \, \theta
\label{eq:pb_ga}
\end{equation}
where $a\triangleq\left( \Omega g_\theta +1 \right)^{-1}$ and
$g_\theta\triangleq c_{M,i}/2\sin^2\theta$. Moreover, using the
expression (3.312.1) in \cite[p.305]{table_integ}, after some
algebraic manipulations, we obtain the following closed form
expression for (\ref{eq:pb_ga}):
\begin{equation}
{\rm Pr}_{b,e}^{K-GA} (K) = \sum_{i=1}^{\Theta_M} C_{M,i} \pi^{-1}
K \int_0^{\frac{\pi}{2}} {\rm B}(K,1+g_\theta \Omega) {\rm
d}\theta
 \label{eq:exPb_ga}
\end{equation}
where ${\rm B}(x,y)\triangleq\int_0^1 t^{x-1} (1-t)^{y-1} {\rm
d}t$ denotes the beta function. For a given $\omega$, it is clear
from (\ref{eq:exPb_ga}) that ${\rm Pr}_{b,e}^{K-GA}(K)$ depends on
both $K$ and $\lambda$. Note that ${\rm B}(x,y)$ decreases
exponentially with $x$ at a given $y$. Therefore, for a given
$\lambda$, the system BER in (\ref{eq:exPb_ga}) decreases
exponentially with respect to $K$ due to improvements in the MUDiv
at the cost of increased $E_T^{K-GA}$ in (\ref{eq:sys_erg}).

\subsection{Upper bound expression on BER}
\label{sec:ubPb_ga} The finite range of the integral in
(\ref{eq:pb_ga}) can be eliminated by considering the minimum
value of $g_\theta$. Thus, substituting $\theta=\pi/2$ in
(\ref{eq:pb_ga}), we find the UB expression for (\ref{eq:exPb_ga})
with $g_u\triangleq{c_{M,i}}/2$ (Chernoff bound) as
\begin{equation}
{\rm Pr}_{b,e}^{K-GA} (K) \leq \sum_{i=1}^{\Theta_M} C_{M,i}
\pi^{-1} K \hspace{0.1cm} {\rm B}(K,1+g_u \Omega) ={\rm
Pr}_{b,u}^{K-GA} (K). \label{eq:ubPb_ga}
\end{equation}
This UB BER is clearly less complex than (\ref{eq:exPb_ga}) since
it does not contain integration.

\subsection{Approximate BER expression}
\label{sec:apPb_ga} Inserting (\ref{eq:pdf_rho_g}) and the
following approximation of (\ref{eq:ber_awgn}) \cite{kote_tw08}:
${\rm Pr}_b(M,\rho)\approx 0.2\hspace{0.1cm} {\rm e}^{-g_a \rho}$,
where $g_a\triangleq1.5 /(M-1)$, into (\ref{eq:pb_rho}), the
approximate BER expression can be written as
\begin{equation}
{\rm Pr}_{b,e}^{K-GA} (K) \approx 0.2 \hspace{0.1cm} K
\hspace{0.1cm} {\rm B}(K, 1+g_a \Omega)={\rm Pr}_{b,a}^{K-GA} (K).
\label{eq:apPb_ga}
\end{equation}
Note that in comparison to the exact and UB BER expressions, which
have multiple summation terms of beta functions, the expression
(\ref{eq:apPb_ga}) requires minimum computations with a single
beta function.

\begin{figure}
\includegraphics[width=12.7cm]{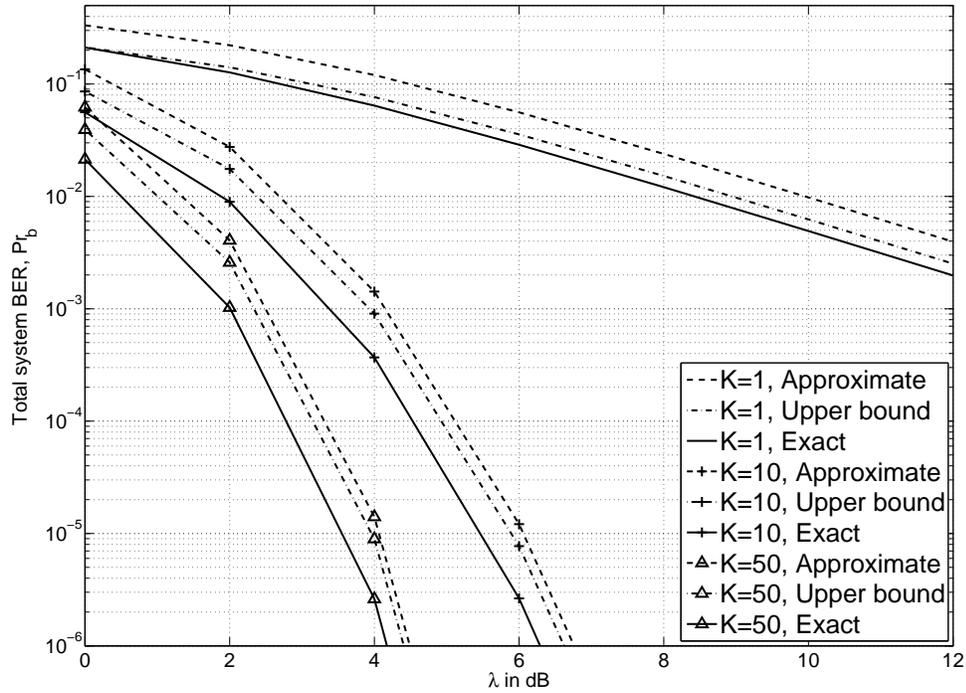}
\caption{Total system BERs ${\rm Pr}_{b,e}(K)$, ${\rm
Pr}_{b,u}(K)$ and ${\rm Pr}_{b,a}(K)$ for $K$-GA when $K=
\bar{K}=1,10,50$.} \label{fig:Pb_ga_ra}
\end{figure}

Fig.~\ref{fig:Pb_ga_ra} illustrates the exact, UB, and approximate
BER's, i.e., (\ref{eq:exPb_ga}), (\ref{eq:ubPb_ga}), and
(\ref{eq:apPb_ga}), of the $K$-GA scheme for $\bar{K} = K = 1, 10,
50$. It can be seen from this figure that the UB and approximate
expressions produce the BER curves which lay within 0.5~dB of the
exact BER.

\section{Optimal selection of the number of active users}
\label{sec:opt} Two scenarios are considered in this section for
selecting the number of active users $K$ optimally: (i) minimizing
${\rm Pr}_b^{K-GA} (\cdot)$ while $E_T^{K-GA}$ remains constant
and (ii) minimizing $E_T^{K-GA}$ while ${\rm Pr}_b^{K-GA} (\cdot)$
is constrained to be acceptably small.\footnote{The exact, UB or
approximate BER's can be considered. We use the notation ${\rm
Pr}_b^{K-GA} (\cdot)$ to refer to any of these three BER
expressions, i.e., ${\rm Pr}_b^{K-GA} (\cdot) \in \{ {\rm
Pr}_{b,e}^{K-GA} (K), \, {\rm Pr}_{b,u}^{K-GA} (K), \, {\rm
Pr}_{b,a}^{K-GA} (K) \} $.} Optimization problems for each
scenario are provided.

The set of candidate values of $K$ is the set of all positive
integers smaller than or equal to $\bar{K}$, i.e, ${\cal K}
\triangleq \{ 1, \cdots, \bar{K} \}$. Note that due to hardware
design limitations ${\cal K}$ can be just a set of some integers
smaller than or equal to $\bar{K}$. The latter case can be easily
adopted in the methods developed further.

\subsection{Optimal selection of $K$ based on system BER minimization}
Using (\ref{eq:txpw_k}) and (\ref{eq:sys_txerg}), under the finite
energy constraint, we can find that the achievable energy gain for
$E_d(K)$ determines $\lambda(K)$ in ${\rm Pr}_b^{K-GA} (\cdot)$ as
follows\footnote{The argument K is added here to emphasize that
$\lambda$ is a function of $K$ in the BER minimization-based
problem.}
\begin{equation}
\label{eq:lambda_K2} \lambda(K) = E_d(K)~c / (N T_s) = \left(
(\bar{K}-K)\alpha^{-1} + 1 \right) E_d^{GA}~c / (N T_s)
\end{equation}
where $c = \bar{K} (\sum_{k=1}^{\bar K} \sigma_k^{-2} )^{-1} /
\omega$. It is worth mentioning that $E_d^{GA} / (N T_s)$ in
(\ref{eq:lambda_K2}) stands for the average transmit power over
$N$ slots when $K=\bar{K}$. Thus, it can be denoted as
$\lambda^{GA}$. Using this notation, (\ref{eq:lambda_K2}) can be
represented as
\begin{equation}
\label{eq:lambda_K} \lambda(K) = \left( (\bar{K}-K)\alpha^{-1} + 1
\right)~c~\lambda^{GA}.
\end{equation}

It can be seen from (\ref{eq:lambda_K}) that a power gain of
$\left( (\bar{K}-K)\alpha^{-1} + 1 \right)c$ in ${\rm Pr}_b^{K-GA}
(\cdot)$ is achieved if $K < \bar{K}$. Denoting $P_T = E/(N T_s)$
as the total {\it average power} consumed by all users during one
slot, $\lambda(K)$ can be also expressed in terms of $P_T$ as
$\lambda(K) = c ( P_T - K \alpha^{-1} \lambda^{GA})$.

Using $\lambda (K)$ in (\ref{eq:lambda_K}), we optimize $K$ for a
given $\omega$ to minimize the system BER while satisfying the
finite energy constraint. The corresponding optimization problem
can be mathematically formulated as
\begin{equation}
K_b^* = \arg\min_{K\in{\cal K}} {\rm Pr}_b^{K-GA} (K)
\hspace{.5cm} \text{subject to} \hspace{0.5cm} E_T^{K-GA} = E.
\label{eq:opt_prob1_minPb}
\end{equation}

One way to solve (\ref{eq:opt_prob1_minPb}) is to employ a binary
search over $K\in{\cal K}$ through direct computation of ${\rm
Pr}_b^{K-GA} (\cdot)$. However, direct evaluation of ${\rm
Pr}_b^{K-GA} (\cdot)$ is computationally complex, and such an
approach can be inaccessible for applications sensitive to high
computational complexity. Therefore, an approach, which avoids
direct computation of ${\rm Pr}_b^{K-GA} (\cdot)$ for all $K$, is
proposed.

To this end, let us relax $K$ to be a real number\footnote{While
relaxing $K$ to be a real number, we also assume that ${\rm
Pr}_b^{K-GA} (\cdot)$ is continuous on ${\cal K}$ and
differentiable at all points on ${\cal K}$.} such that $K \in [1,
\bar{K}]$. Let us also define $\eta(\cdot) \triangleq
\frac{\partial}{\partial K} {\rm Pr}_b^{K-GA} (\cdot)$. It can be
observed that ${\rm Pr}_b^{K-GA} (\cdot)$ is convex with respect
to $K$ due to the fact that $\frac{\partial^2}{\partial^2 K}{\rm
Pr}_b^{K-GA} (\cdot)\geq0$. Thus, the minimum of ${\rm
Pr}_b^{K-GA} (\cdot)$ over $K \in [1, \bar{K}]$ can be found by
minimizing $\big| \eta(\cdot) \big|$ for a given normalized
$\Omega$, i.e., $\Omega_N = \omega \lambda^{GA}$. Note that this
minimum is unique. Therefore, denoting $K^*$ as a real-valued
solution, the corresponding optimal solution can be given by
\begin{equation}
K^*=\arg\min_{K\in[K_n,K_1] } \Big| \eta (K) \Big|.
\label{eq:sol_optKb}
\end{equation}

Recall that ${\cal K}$ is a finite set of integers, and the
optimal $K^*_b \in {\cal K}$ may not be equivalent to $K^*$ in
(\ref{eq:sol_optKb}). Therefore, (\ref{eq:sol_optKb}) should be
reformulated as
\begin{equation}
K^*_b =
\begin{cases}
\bar{K}, & \text{if} \hspace{0.1cm} \text{sign}(\eta(K=\bar{K}))= -1\\
1, & \text{else if} \hspace{0.1cm} \text{sign}(\eta(K=1)) = 1\\
\arg\min_{ K\in\{K_i,K_{i+1}\} } {\rm Pr}_b^{K-GA} (K), &
\text{otherwise}
\end{cases}
\label{eq:opt_prob1-2_minPb}
\end{equation}
where ${\rm sign}(a)=|a|/a$ for $a\in\mathbb{R}$ with ${\rm
sign}(0)=1$, and $K_i\in{\cal K}$ is the largest integer smaller
than $K^*$ that satisfies the equality ${\rm sign} (\eta(K_i))
{\rm sign}(\eta(K_{i+1})) = -1$.

In order to find $K^*_b\in{\cal K}$, we first compute ${\rm
sign}(\eta(K))$ at $K=\bar{K}$ (and/or $K=1$). If the resulting
${\rm sign}(\cdot)$ is $-1$ (or $1$), then we select $K^*_b =
\bar{K}~(\text{or } 1).$ Otherwise, $K_i \in {\cal K}$ can be
found by binary search algorithm followed by selecting $K^*_b$ at
whichever of $K_i$ or $K_{i+1}$ that has a smaller ${\rm
Pr}_b^{K-GA} (\cdot)$.

Considering, for example, the case when ${\rm Pr}_b^{K-GA} (\cdot)={\rm
Pr}_{b,a}^{K-GA} (K)$, it is shown in Appendix that
\begin{equation}
\eta (K) = \frac{1}{K} - \sum_{l=0}^{f(K)}\frac{1}{K+l} -
f'(K)\sum_{l=0}^{K-1}\frac{1}{1+f(K)+l} \label{eq:sol_opt_prob1}
\end{equation}
where $f(K)\triangleq g_a\omega \lambda$.

\begin{figure}
\includegraphics[width=12.7cm]{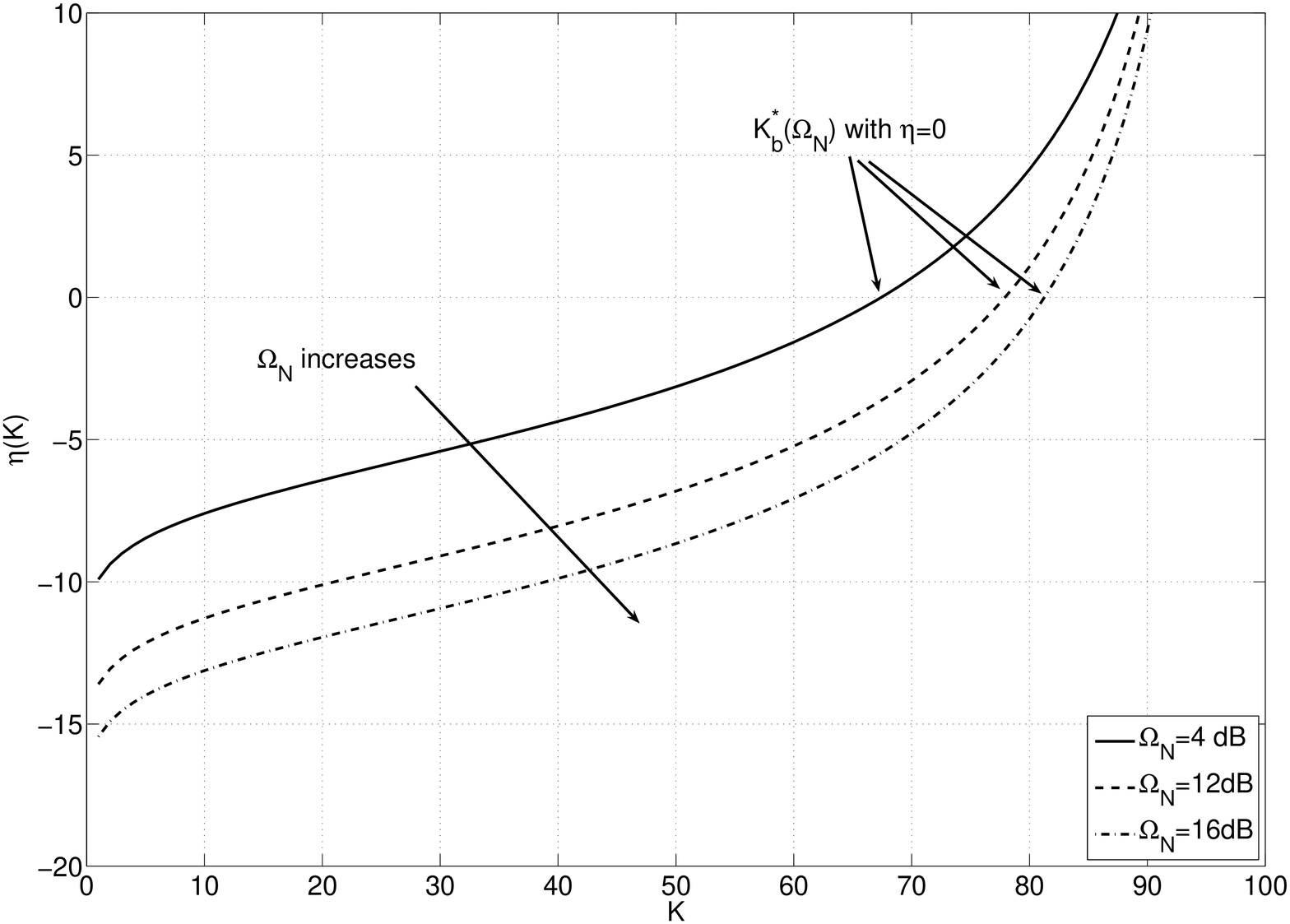}
\caption{Function $\eta(\cdot)$ versus $K$ when $\bar{K}=100,$
$M=4,$ and $\alpha=1.$} \label{fig:eta_vs_K}
\end{figure}

Given $\Omega_N$ and a finite set ${\cal K}=\{1, 2, \cdots,
\bar{K} \},$ $\eta (K)$ is illustrated in Fig. \ref{fig:eta_vs_K}.
It is shown that neither $K\rightarrow1$ nor $K\rightarrow\bar{K}$
may minimize ${\rm Pr}_b^{K-GA} (\cdot)$. Therefore, for a given
$\Omega_N$, there exists an optimal $1 \leq K^*_b\leq\bar{K}$
minimizing ${\rm Pr}_b^{K-GA} (\cdot)$. For example, it can be seen
from the figure that when $M=4$ and $\bar{K}=100$, $K^*_b\approx
67$ minimizes ${\rm Pr}_b^{K-GA}(\cdot)$ at $\Omega_N=4$~dB.

\subsection{Optimal selection of $K$ based on $E_T^{K-GA}$ minimization}
\label{sec:opt_K_minPT} If an MU system is capable of recovering
properly the transmitted information as long as the system BER is
less than or equal to a predefined desired level, the optimal $K$
can be found via minimization of $E_T^{K-GA}$ in
(\ref{eq:sys_erg}) subject to the constraint ${\rm Pr}_b^{K-GA}
(\cdot) \leq {\rm BER}_t$ where ${\rm BER}_t$ is the required
target BER. In this problem, different from the previous problem,
$E_T^{K-GA}$ is the optimization variable, while $\lambda$
(equivalently $\Omega_N=\Omega$) is fixed.

Two cases of delay tolerant and delay sensitive systems are of
interest.

\subsubsection{Delay tolerant (DT) systems} The constrained optimization
problem for finding optimal $K$ can be written in this case as
\begin{equation}
K_{dt}^* = {\rm arg}\min_{K\in{\cal K}} E_T^{K-GA} \hspace{0.5cm}
\text{subject to} \hspace{0.5cm} {\rm Pr}_b^{K-GA} (K) \leq {\rm
BER}_t. \label{eq:ac}
\end{equation}
Note that for a given $\Omega$, ${\rm Pr}_b^{K-GA} (\cdot)$ and
$E_T^{K-GA}$ are, respectively, monotonically decreasing and
monotonically increasing functions of $K$. Thus, among all values
of $K$ satisfying the constraint ${\rm Pr}_b^{K-GA} (\cdot) \leq
{\rm BER}_t$, the smallest $K\in{\cal K}$ which minimizes
$E_T^{K-GA}$ is the solution of (\ref{eq:ac}).

In order to find $K_{dt}^*\in{\cal K}$, we first need to find
${\rm Pr}_b^{K-GA} (\cdot)$ when $K=\bar{K}.$ If $\bar{K}$ does not satisfy the
system BER requirements, then $K^*_{dt}=0$. When $K^*_{dt}=0$, the
system may allow delays to prevent the waste of the total energy
of all users. Otherwise, the smallest $K \leq \bar{K}$, which
satisfies the system BER requirements, can be searched efficiently
using, for example, a binary search algorithm.

\subsubsection{Delay sensitive (DS) systems} DS systems allow to transmit
data even if ${\rm Pr}_b^{K-GA} (\cdot)~>~{\rm BER}_t$ when $K=\bar{K}.$ Then,
the corresponding constrained optimization problem can be written
as
\begin{equation}
K_{ds}^* =
\begin{cases}
\arg\min_{K\in{\cal K}} \hspace{0.1cm} E_T^{K-GA} \; \text{subject
to} \; {\rm Pr}_b^{K-GA} (K) \leq {\rm BER}_t, & {\rm if} \; {\rm
Pr}_b^{K-GA} (\bar{K}) \!\leq\! {\rm BER}_t \\
K_1, & {\rm otherwise}.
\end{cases}
\label{eq:ac_2}
\end{equation}
The problem (\ref{eq:ac_2}) can be solved similar to the previous
one. The only difference is that $K^*_{ds}=\bar{K}$ even if
$K=\bar{K}$ is not sufficient to satisfy the constraint ${\rm
Pr}_b^{K-GA} (\cdot) \leq {\rm BER}_t$.

\section{Asymptotic analysis}
\label{sec:analysis_K}

\subsection{Asymptotic analysis of optimal $K$ based on $E_T^{K-GA}$
minimization} In general, the optimum $K$ based on $E_T^{K-GA}$
minimization under fixed system BER cannot be found in closed
form. However, its asymptotic behavior can be studied
analytically. Recall that the derived system BER expressions
depend on the beta function ${\rm B}(\cdot,\cdot)$. Thus, we first
study the asymptotic behavior of ${\rm B}(\cdot,\cdot)$ with
respect to $K$.

The following theorem summarizes the asymptotic behavior of the
beta function.
\begin{thry}
{\it Let $x$ and $y$ be two positive integers. When
$x\rightarrow\infty$, we have}
\begin{equation}
\lim_{x\rightarrow\infty}x^y \hspace{0.1cm} {\rm B}(x,y)=
\Gamma(y) \label{eq:bigO_beta}
\end{equation}
{\it where $\Gamma(y)=\int_0^{\infty}t^{y-1}{\rm e}^{-t}{\rm d}t$
denotes the complete gamma function.} \label{thr1}
\end{thry}
\emph{Proof:} The beta function can be alternatively represented
in terms of the following ratio of complete gamma functions
\begin{equation}
{\rm B}(x,y)=\frac{\Gamma(x)\Gamma(y)}{\Gamma(x+y)}.
\label{eq:alt_beta}
\end{equation}
Using the expression $\Gamma(x)=(x-1)!$ for the gamma function, we
can find the following ratio
\begin{equation}
\frac{\Gamma(x)}{\Gamma(x+y)}=\frac{(x-1)!}{(x+y-1)!}=
\frac{1}{(x+y-1)(x+y-2)\cdots(x+1)x}.
\label{eq:ratio_gamma}
\end{equation}
Since $(x+y-1)(x+y-2)\cdots(x+1)x$ in (\ref{eq:ratio_gamma}) is
dominated by the first power term $x^y$, the ratio in
(\ref{eq:ratio_gamma}), for $x\rightarrow\infty$, becomes
\begin{equation}
\lim_{x\rightarrow\infty}\Gamma(x)/\Gamma(x+y) = x^{-y} .
\label{eq:approx_ratio_gamma}
\end{equation}
Thus, when $x\rightarrow\infty$, inserting
(\ref{eq:approx_ratio_gamma}) into (\ref{eq:alt_beta}) reveals the
asymptotic behavior of (\ref{eq:alt_beta}) as
\begin{equation}
\lim_{x\rightarrow\infty} \hspace{0.1cm} {\rm B}(x,y) =
\lim_{x\rightarrow\infty} \frac{\Gamma(x)}{\Gamma(x+y)} \Gamma(y)
= x^{-y} \hspace{0.1cm} \Gamma(y). \label{eq:asymp_beta}
\end{equation}
Since for a given $y$, $\Gamma(y)$ in (\ref{eq:asymp_beta}) is
fixed, (\ref{eq:bigO_beta}) is obtained when $x\rightarrow\infty$.
\hfill $\blacksquare$

Theorem~\ref{thr1} enables us to evaluate the system BER for two
asymptotic cases of (i) large $K$ and (ii) high SNR. We also aim
at investigating how the optimal $K$ scales asymptotically. For
simplicity, only the approximate BER expression (\ref{eq:apPb_ga})
is considered in the further analysis.

{\it For the case of large $K$}, we aim at analyzing ${\rm
Pr}_{b,a}^{K-GA} (K)$ versus $K$ and SNR. As per Theorem \ref{thr1},
for large values of $x$, the following approximation holds true
${\rm B}(x,y)\approx x^{-y}\Gamma(y)$. Then, when interpreting $x$
and $y$ in (\ref{eq:bigO_beta}) as $K$ and $1+g_a\Omega$,
respectively, the system BER can be expressed for large values of
$K$ as
\begin{equation}
\begin{split}
{\rm Pr}_{b,a}^{K-GA} (K) & = K^{-g_a\Omega} 0.2
\hspace{0.1cm} \Gamma(1+g_a\Omega)\\
\quad &= \Theta \left( K^{-{\rm SNR}} \right)
\end{split}
\label{eq:asymp_Prb}
\end{equation}
where we use the alternative notation SNR$=\Omega$ in the last
expression. Therefore, (\ref{eq:asymp_Prb}) shows how the system
BER scales with respect to the MUDiv gain if $K$ is large.

{\it For another asymptotic case of large SNR}, ${\rm
Pr}_{b,a}^{K-GA}(K)$ can also be expressed in terms of SNR and
$K$. Specifically, using the fact that ${\rm B}(x,y)={\rm
B}(y,x),$ it follows straightforwardly from Theorem \ref{thr1}
that for large $y$, ${\rm B}(x,y)\approx y^{-x}\Gamma(x).$
Therefore, the system BER ${\rm Pr}_{b,a}^{K-GA} (K)$ can be expressed
for large SNR as
\begin{equation}
\label{eq:asymp_Prb2}
\begin{split}
{\rm Pr}_{b,a}^{K-GA} (K) &= \Omega^{-K} 0.2
\hspace{0.1cm} g_a^{-K} \Gamma(K+1) \hspace{0.1cm} \\
\quad &= \Theta \left( {\rm SNR}^{-K} \right).
\end{split}
\end{equation}
It follows from (\ref{eq:asymp_Prb2}) that the total system BER
scales inversely with the order of the SNR, i.e, the MUDiv is
equal to $K$.

Using (\ref{eq:asymp_Prb}) and (\ref{eq:asymp_Prb2}), we can find
the optimal $K^* \in {\cal K}$, i.e., either $K^*_{dt}$ for the
delay tolerant or $K^*_{ds}$ for the delay sensitive systems.

{\it In the case when $K$ is large}, it can be found using
(\ref{eq:asymp_Prb}) that for given $\omega$, $\lambda$, and ${\rm
BER}_t$, the optimal $K^* \in{\cal K}$, i.e., $K^*_{dt}$ for the
delay tolerant or $K^*_{ds}$ for the delay sensitive systems, must
satisfy the following inequality
\begin{equation}
\begin{split}
K^* &\geq \left(0.2\Gamma(1+g_a\Omega) \right)^{\frac{1}
{g_a\Omega}} {\rm BER}_t^{-\frac{1}{g_a\Omega}} \\
\quad &= \Theta \left({\rm BER}_t^{-1/{\rm SNR}} \right).
\end{split}
\label{eq:asymp_OptK_minPT}
\end{equation}
It follows from (\ref{eq:asymp_OptK_minPT}) that for large $K$ and
a given SNR, $K^*$ is an exponentially decreasing function of
${\rm BER}_t$. 

{\it In the case of high SNR}, it can be found from
(\ref{eq:asymp_Prb2}) that the optimal $K^*$, which guarantees
that the target BER is archived, i.e., the constraint ${\rm
Pr}_{b,a}^{K-GA} (K)\leq {\rm BER}_t$ is satisfied, must obey
the following inequality
\begin{equation}
\label{eq:asymp_OptK_minPT2} K^* \geq \frac{\log {\rm
BER}_t^{-1}}{\log g_a\Omega} = \Theta \left( \frac{ \log {\rm
BER}_t^{-1} }{ \log {\rm SNR} } \right).
\end{equation}
For a given ${\rm BER}_t$, it follows from
(\ref{eq:asymp_OptK_minPT2}) that the corresponding optimal $K^*$,
i.e., $K^*_{dt}$ for the delay tolerant or $K^*_{ds}$ for the
delay sensitive systems, is proportional to the inverse of
$\log{\rm SNR}$. Moreover, unlike the case of large $K$, in the
case of high SNR, $K^*$ decreases in a log-scale with ${\rm
BER}_t$.

\subsection{Asymptotic analysis of optimal $K$ based on the
system BER minimization} \label{behavior_opt_K:sec} We again
consider two cases of (i) large $K$  and (ii) high SNR and study
the asymptotic behavior of optimal $K^*_b$, i.e., we study the
asymptotic behavior of the solution of the optimization problem
(\ref{eq:asymp_Prb2}). For simplicity, but without any loss of
generality, we assume that $c= 
1$.

In the case when $K$ is large, we first determine how the system
BER scales with $K$ while satisfying the finite energy constraint.
The corresponding power gain given by (\ref{eq:lambda_K}) is $G_p
\triangleq \left( (\bar{K}-K)\alpha^{-1} + 1 \right).$ Using this
notation and (\ref{eq:asymp_Prb}), the system BER can be
asymptotically expressed as
\begin{equation}
\label{eq:asymp_Prb3}
\begin{split}
{\rm Pr}_{b,a}^{K-GA} (K) &=  0.2\Gamma(1 +
g_a G_p \Omega_N) K^{- g_a G_p \Omega_N} \\
\quad &= \Theta \left( {K}^{-G_p {\rm SNR}} \right)
\end{split}
\end{equation}
where we use the alternative notation SNR$=\Omega_N$. Therefore,
if $K \gg \bar{K} - K$, the achievable MUDiv gain is determined by
$G_p(K) \cdot {\rm SNR} = \left( (\bar{K}-K)\alpha^{-1} + 1
\right) \cdot {\rm SNR}$ instead of ${\rm SNR}$. It is also worth
mentioning that, for a given SNR, the asymptotic system BER scales
exponentially with $K^{(\bar{K}-K)/\alpha}$. The latter means, in
particular, that the achievable system BER is lower in the case of
using optimal $K$ as compared to the case when all users are
active, i.e., $K=\bar{K}$.

In the case of high SNR, using $G_p$ and (\ref{eq:asymp_Prb2}),
the asymptotic expression for the system BER can be obtained as
\begin{equation}
\label{eq:asymp_Prb4}
\begin{split}
{\rm Pr}_{b,a}^{K-GA} (K) &= 0.2 g_a^{-K} \Gamma(K+1) (G_p
\Omega_N)^{-K} \\
\quad &= \Theta \left( (G_p {\rm SNR})^{-K}
\right).
\end{split}
\end{equation}
It follows from (\ref{eq:asymp_Prb4}) that the asymptotic system
BER benefits from the MUDiv power gain $G_p^{-K} = \left(
(\bar{K}-K)\alpha^{-1} + 1 \right)^{-K}$ at the cost of having the
diversity order $K < {\bar K}$.

Finally, inserting (\ref{eq:asymp_Prb4}) into
(\ref{eq:sol_optKb}), we obtain that
\begin{equation}
\label{eq:optKb_limit}
\begin{split}
\lim_{SNR\rightarrow\infty}K^*_b &= \arg \min_K \left|
\frac{\partial}{\partial K} (\ref{eq:asymp_Prb4}) \right|
= \arg \min_K \left| G_p \cdot {\rm SNR} \right|\\
\quad & = \bar{K} + \alpha
\end{split}
\end{equation}
where $0 \leq \alpha \leq 1$. Since optimal MUDiv $K$ is
restricted to be integer, it can be concluded from
(\ref{eq:optKb_limit}) that the MUDiv $K = \bar{K}$ is optimal
when SNR$\rightarrow\infty$. The latter means that the maximum
available MUDiv should be used for energy unlimited systems that
agrees with known results.

\section{Extension to multiple antenna MU systems}
\label{sec:ext_mu} The optimization problems proposed in
Section~\ref{sec:opt} can be extended to multiple antenna MU
systems, which will also allow to use the benefits of multiple
antenna techniques \cite{tse_fundamental,dai_jsp07}. Toward this
end, a generalized expression for the average BER has to be
derived. For brevity, we consider only the approximate average
system BER case.

Let $D$ denote the multiple antenna diversity order. Then, in the
multiple antenna case, the degrees of freedom (DOF) of $y$ in
(\ref{eq:pb_rho}) extends to $2D$, that is, $y\sim\chi_{2D}^2$
where $\chi^2_{2D}$ stands for the Chi-squared distribution with
$2D$ DOF (refer also to \cite{dai_jsp07}). Therefore, for given
$K$ and $D$, the expression for $p_\rho^{K-GA}(y)$ in
(\ref{eq:pdf_rho_g}) can be generalized as \cite{tse_fundamental}
\begin{equation}
p_\rho^{K-GA}(y)=K\frac{{\rm e}^{-y/\Omega}} {\Omega}\gamma \left(
D,y/\Omega \right)^{K-1} \frac{\left( y/\Omega \right)^{D-1}}
{\Gamma(D)^K} \label{eq:pdf_rho_mamu}
\end{equation}
where $\gamma(a,b)\triangleq\int_0^b t^{a-1}{\rm e}^{-t} {\rm d}t$
denotes the lower incomplete gamma function \cite{table_integ}.

Inserting (\ref{eq:pdf_rho_mamu}) into (\ref{eq:pb_rho}), we
obtain the average BER in the multiple antenna case as
\begin{equation}
{\rm Pr}^{K-GA}_b (K) =0.2 K \int_{0}^{\infty} t^{D-1} {\rm
e}^{-(1+g_a\Omega)t} \gamma(D,t)^{K-1}/\Gamma(D)^K \hspace{0.1cm}
{\rm d}t. \label{eq:aprx_prb_n}
\end{equation}
Finally, the optimization problems proposed Section~\ref{sec:opt}
can be straightforwardly extended to the case of multiple antenna MU
systems by using (\ref{eq:aprx_prb_n}) instead of the
corresponding BER expressions for the single antenna case. As an
example, an extension of the problem (\ref{eq:opt_prob1_minPb}) to
the case of multiple antenna MU systems will be investigated
numerically in the following section.

\section{Numerical results}
\label{sec:num} Consider an MU system with Gray-coded square M-QAM
of size $M \in \{ 4, 64 \}$. Let $(C_{M,1}, c_{M,1})$ in
(\ref{eq:ber_awgn}) be ($1,1$) for $M=4$, while
$\{(C_{M,1},c_{M,1}),\cdots,$ $(C_{M,5},c_{M,5})\}$ be $\{(7/12,
1/21), (1/2, 3/7),$\\
$(-1/12, 25/21), (1/12, 9^2/21), (-1/12, 13^2/21)\}$ for $M=64$ \cite{chha03}.
Independent log--normal distributed shadowing with mean $\mu=1$
and standard deviation $\nu=5$ is assumed with pathloss 0~dB.
Considering the approximate BER, i.e., ${\rm Pr}_b^{K-GA} (\cdot) = {\rm
Pr}_{b,a}^{K-GA} (K)$, the optimal $K$, i.e., $K^*_b$ of
(\ref{eq:opt_prob1-2_minPb}), $K^*_{dt}$ of (\ref{eq:ac}), or
$K^*_{ds}$ of (\ref{eq:ac_2}), are found. The set ${\cal K}=\{1,
2, \cdots, \bar{K}\}$, $c=1$, $\alpha \in \{1, 2, 7.81, 31.25 \},$
and $E_d(K)/E_f \in [\alpha, \bar{K} - 1 + \alpha]$ are
used. Note that the parameter $\alpha=2$ corresponds to
the standard case when $280$ pilot sub-carriers and $560$ data
sub-carriers are used per one sub-channel in $10$~MHz uplink WiMAX
(IEEE 802.16e) \cite{80216e_2006}. For comparisons, we also
consider other parameter values. For example, the parameter
$\alpha=7.81$ can be obtained by using 32 pilot and 250 data
sub-carriers while $\alpha=31.25$ results from using 32 pilot
and 1000 data sub-carriers.

In the case when $\bar{K} = 100$ and $\alpha = 1$, the ratio
$\lambda(K) / \lambda^{GA}$ (or, equivalently, $E_d(K) /
E_d^{GA}$) is set at the values between 0~dB and 20~dB depending
on $K$. The generic GA scheme is also depicted for comparison.

\subsection{Minimizing the system BER}
{\it Example 1:} In our first example, we consider the problem
(\ref{eq:opt_prob1-2_minPb}) and the case when for a given
$\Omega_N \in \{5,10\}$~dB and $\alpha=2$, total energy grows with
$\bar{K}$. Note that $E_T^{K-GA}$ (or, equivalently, the average
power $P_T (K)$) is an increasing function of $\bar{K}$.

\begin{figure}
\includegraphics[width=12.7cm]{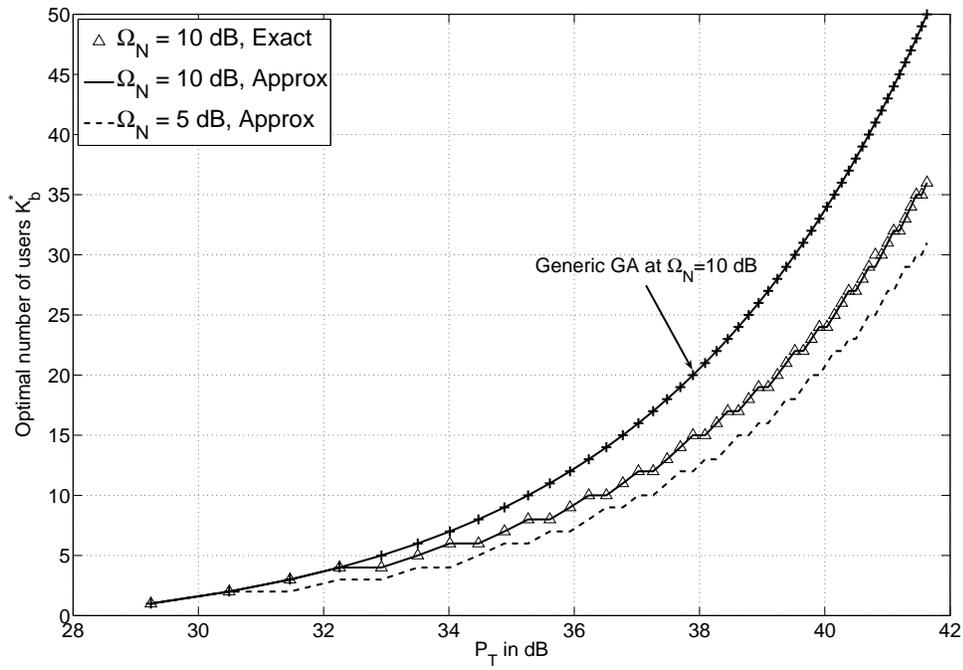}
\caption{Optimal number of users $K^*_b$ versus $P_T$ for
different $\Omega_N$, $M=4,$ and $\alpha=2.$}
\label{fig:OptK_vs_PT}
\end{figure}

Fig.~\ref{fig:OptK_vs_PT} shows $K^*_b$ of
(\ref{eq:opt_prob1-2_minPb}) versus $P_T$ for various values of
$\Omega_N$. It can be seen from the figure that $K^*_b$ increases
with respect to $P_T$. The latter means that the maximum available
MUDiv should be used for energy unlimited systems, while the
optimal MUDiv can be significantly smaller than the maximum
available MUDiv $\bar{K}$ for energy limited systems. It can also
be observed that for a given $P_T$ and low $\Omega_N$, the optimal
MUDiv $K^*_b$ is also small and more energy should be allocated
for actual data transmission $E_d (K)$ in order to achieve better
BER. Finally, it can be also seen in this figure that $K^*_b$ of
(\ref{eq:opt_prob1-2_minPb}) that minimizes the approximate BER
coincides with $K^*_b$ of (\ref{eq:opt_prob1-2_minPb}) that minimizes the exact BER, which validates the use of approximate BER.

\begin{figure}
\includegraphics[width=12.7cm]{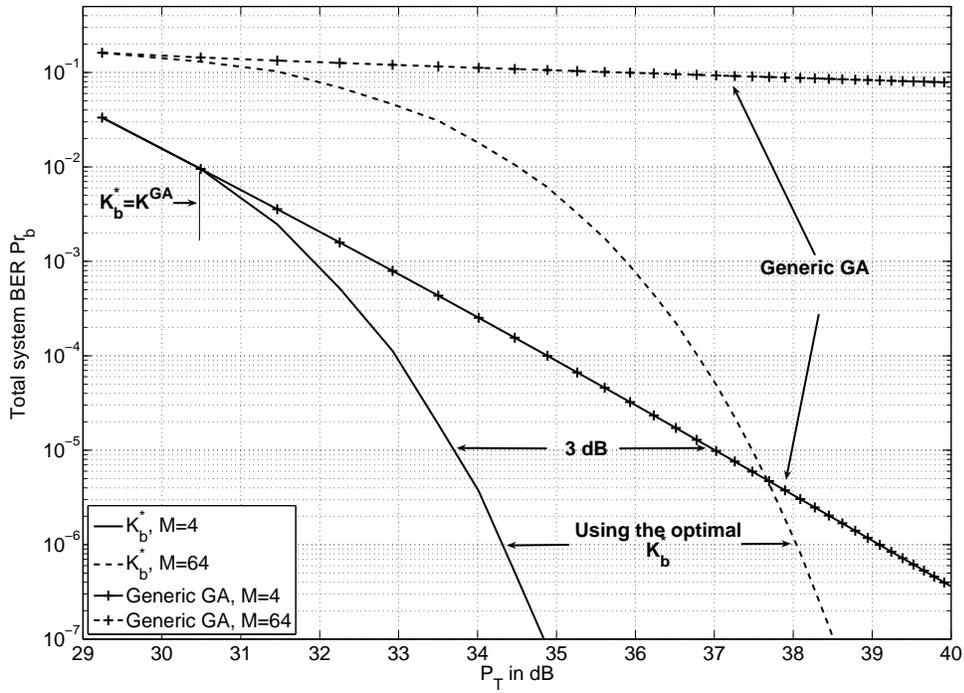}
\caption{Total system BER using $K^*_b$ versus $P_T$ when
$\Omega_N=5$~dB, $M=4,64,$ and $\alpha=2.$} \label{fig:Pb_vs_PT}
\end{figure}

Fig.~\ref{fig:Pb_vs_PT} illustrates the impact of $K^*_b$ on ${\rm
Pr}_{b}^{K-GA} (\cdot)$ versus $P_T$. In this figure,
$\Omega_N=5$~dB and $\alpha=2$ are taken. It can be seen from the
figure that ${\rm Pr}_{b}^{K-GA}(\cdot)$ based on $K^*_b$ is a
decreasing function of $P_T$. Moreover, for $M=4$ and
$P_T\leq30.5$~dB, the generic GA scheme is optimal since it
provides minimum ${\rm Pr}_{b}^{K-GA}(\cdot)$ and
$K^*_b=\bar{K}$, in this case. However, when $P_T\geq 30.5$~dB,
the optimal $K^*_b < \bar{K}$ is obtained. For example, when
$M=4$, $K^*_b$ provides 3~dB power gain at ${\rm Pr}_b=10^{-5}$ as
compared to the generic GA.

{\it Example 2:} In the second example, we consider the problem
(\ref{eq:opt_prob1-2_minPb}) and the case when for a given maximum
achievable MUDiv $\bar{K}$, $E_T^{K-GA}$ grows with $\Omega_N$. In
this case, $\alpha\in\{7.8125, 31.25\}$ and $\bar{K}=50$ are used.

\begin{figure}
\includegraphics[width=12.7cm]{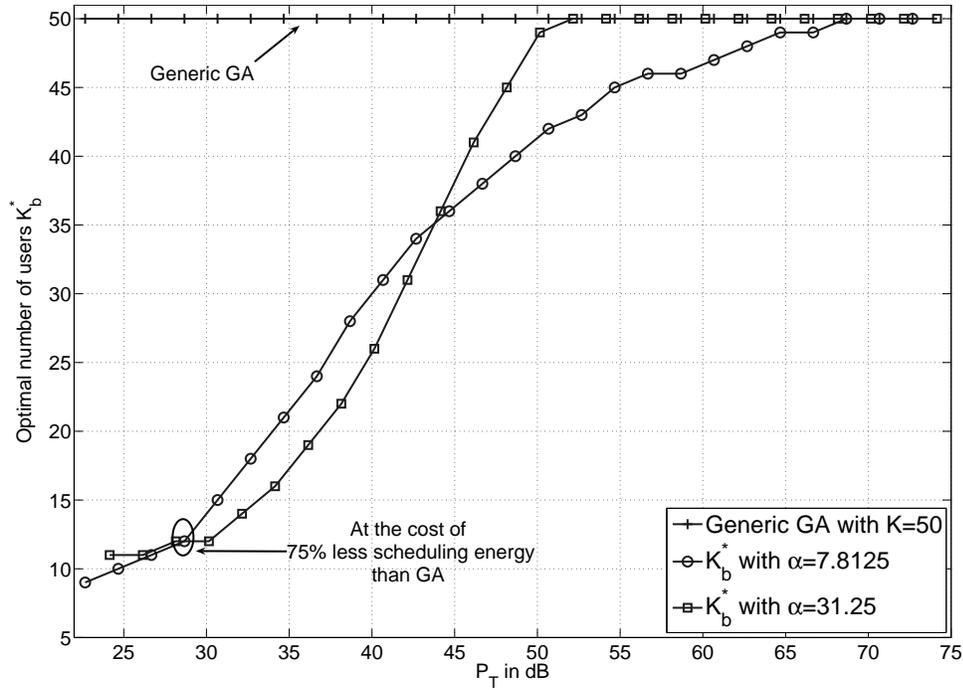}
\caption{Optimal number of users $K^*_b$ versus $P_T$ when
$\bar{K}=50,$ $M=4,$ $\alpha\in\{7.8125,31.25\}.$}
\label{fig:optK_alps}
\end{figure}

Fig.~\ref{fig:optK_alps} shows $K^*_b$ versus $P_T$. It can be
seen from the figure that $K^*_b$ is an increasing function of
$P_T$ and it converges to $\bar{K}$ if more power (energy) is
available for all users in the system. The convergence rate
depends on $\alpha$ and it is higher for larger $\alpha$ and
slower for smaller $\alpha$. Note that the practical values of
$\alpha$ are smaller than both values tested in this example (see
Example~1). It can also be observed that for low $P_T$, less
$K^*_b E_f$ is required to achieve a better system BER than the
one achieved if all $\bar K$ users are active. For example, for
$P_T=28$~dB and $\alpha=7.8125$, the achieved $K^*_b E_f$ for
$K^*_b=12$ is significantly smaller than the one for the
generic GA.

\begin{figure}
\includegraphics[width=12.7cm]{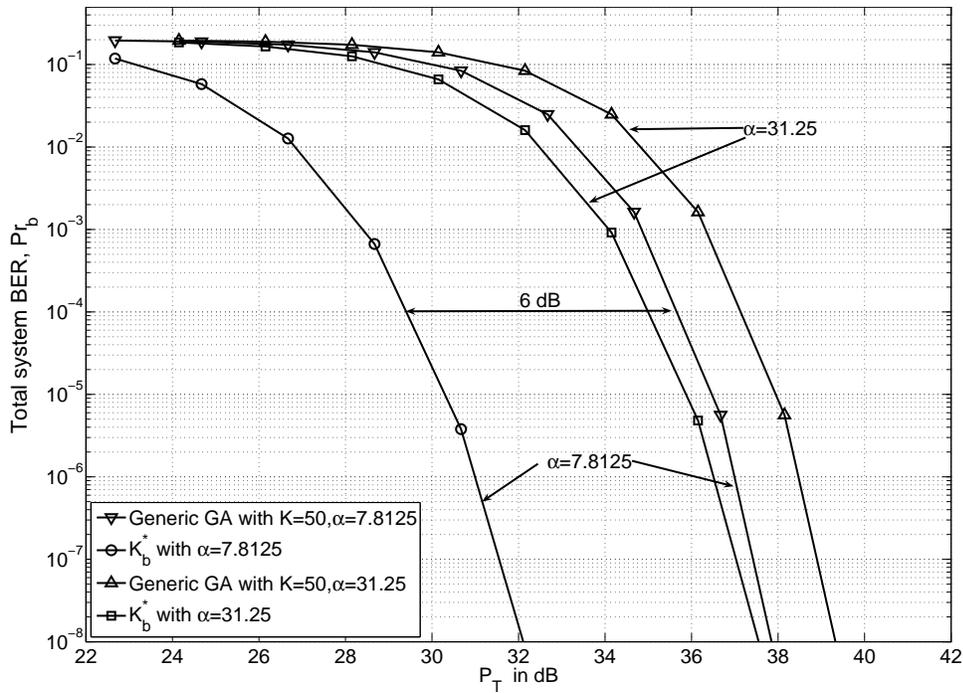}
\caption{Total system BER using $K^*_b$ versus $P_T$ when
$\bar{K}=50,$ $M=4,$ $\alpha\in\{7.8125,31.25\}.$}
\label{fig:Prb_alps}
\end{figure}

In Fig.~\ref{fig:Prb_alps}, the impact of $K^*_b$ on ${\rm
Pr}_b^{K-GA} (\cdot)$ is illustrated versus $P_T$. A significant
power gain is provided by the proposed method as compared to the
generic GA scheme. For example, in the case when $\alpha=7.8125,$
the use of $K^*_b$ provides $6$~dB power gain at ${\rm
Pr}_b^{K-GA} (\cdot)=10^{-4}$. A significant power gain can be observed
even for large $\alpha$, i.e., $\alpha=31.25$. However, regardless
of $\alpha$, the aforementioned power gain vanishes and $K^*_b$
converges to $\bar{K}$ if $P_T \rightarrow \infty$ (see also
Fig.~\ref{fig:optK_alps}).

\subsection{Minimizing the total energy of all users in the system}
{\it Example 3:} In the last example, we consider the problems
(\ref{eq:ac}) and (\ref{eq:ac_2}) for the DT and DS MU systems,
correspondingly. The proposed $K$-GA scheduling scheme based on
$K^*_{dt}$ of (\ref{eq:ac}) and $K^*_{ds}$ of (\ref{eq:ac_2}) is
compared to the generic GA scheduling scheme.

\begin{figure}
\includegraphics[width=12.7cm]{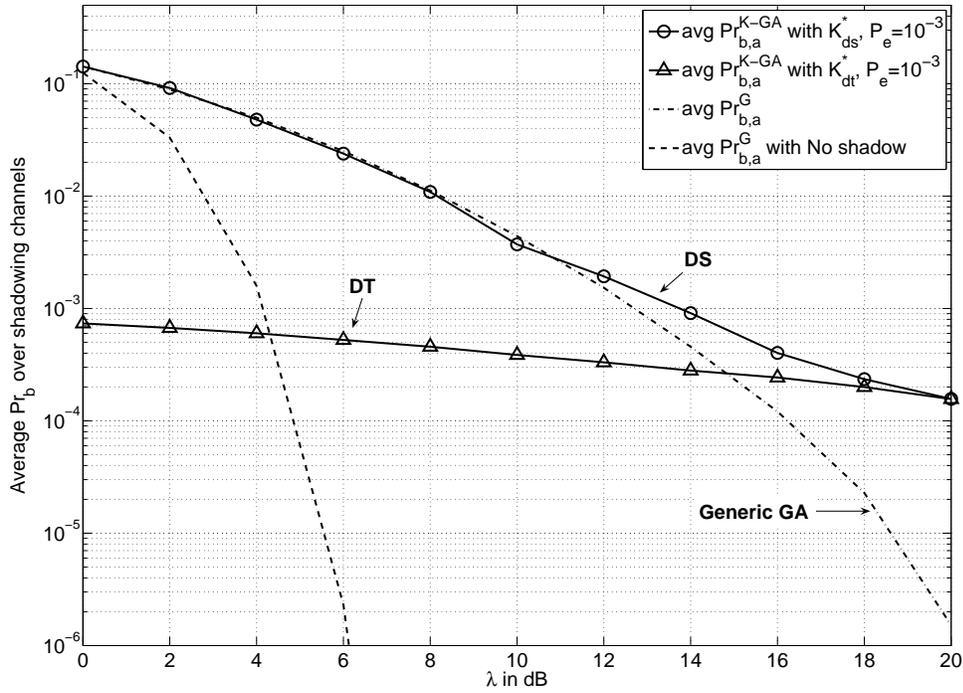}
\caption{Average ${\rm Pr}_b$ of $K$-GA in DS MU and DT MU systems
when $\mu=1,$ $\nu=5,$ ${\rm BER}_t=10^{-3},$ $\alpha=1,$
$\bar{K}=100.$} \label{fig:avg_Pb_aga}
\end{figure}

Fig.~\ref{fig:avg_Pb_aga} shows the error probability of the
proposed $K$-GA scheduling scheme ${\rm Pr}_b^{K-GA}(\cdot)$
averaged over variations of the channel mean $\omega$ versus
$\lambda$. The parameters $\bar{K}=100$ and ${\rm BER}_t=10^{-3}$
are taken. The average error probability of the generic GA is
computed for two cases with and without variations of the channel
mean $\omega$. In can be seen from the figure that the average
${\rm Pr}_b^{K-GA} (\cdot )$ is maintained below the system
requirements (i.e., ${\rm BER}_t=10^{-3}$) for the DT MU system.
For the DS MU system, the average ${\rm Pr}_b^{K-GA}(\cdot )$
is close to the average ${\rm Pr}_{b,a}^{GA} (\cdot )$ at low SNRs
since in order to guarantee a given target BER ${\rm BER}_t$, the
outage is not allowed even if $\bar{K}$ is not sufficiently large.
It can be also seen that as $\lambda$ increases, the DS MU system
performs closer to the DT MU system. It is because $K\leq\bar{K}$
is sufficiently large to guarantee the target ${\rm BER}_t$ in
both cases.

\begin{figure}
\includegraphics[width=12.7cm]{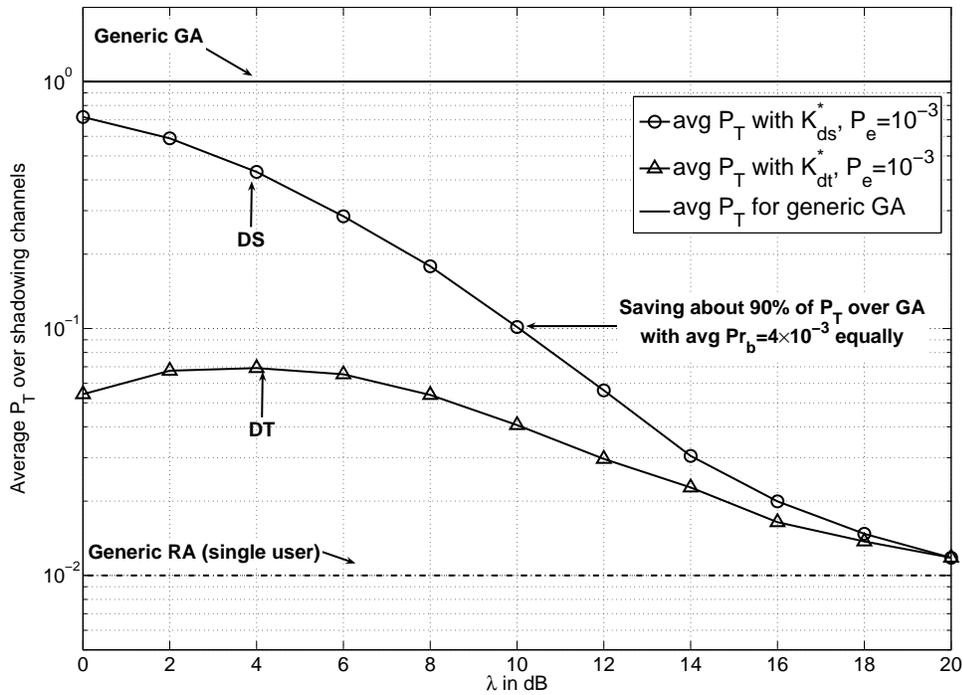}
\caption{Average $P_T$ of $K$-GA in DS MU and DT MU systems when
$\mu=1,$ $\nu=5,$ $\bar{K}=100,$ $\alpha=1,$ ${\rm
BER}_t=10^{-3}.$} \label{fig:avg_PT_aga}
\end{figure}

Based on $K^*_{dt}$ and $K^*_{ds}$, it can also be seen in
Fig.~\ref{fig:avg_PT_aga} that the average $P_T$ normalized by the
power required for the generic GA is a decreasing function of
$\lambda$ for both the DT and DS MU systems. Moreover, the DT MU
system requires less power (energy) than the DS MU system while
satisfying the system requirement on target BER ${\rm BER}_t$. For
example, at $\lambda\approx 10$~dB, the DS MU systems with ${\rm
BER}_t=10^{-3}$ achieves a power saving gain of $10$~dB over the
generic GA for the same average ${\rm Pr}_b=4\times 10^{-3}$ (see
Figs.~\ref{fig:avg_Pb_aga}~and~\ref{fig:avg_PT_aga}).
Figs.~\ref{fig:avg_Pb_aga}~and~\ref{fig:avg_PT_aga} also depict
that as $\lambda$ increases, $P_T$ converges to the power required
for the RA scheduling scheme.

\begin{figure}
\includegraphics[width=12.7cm]{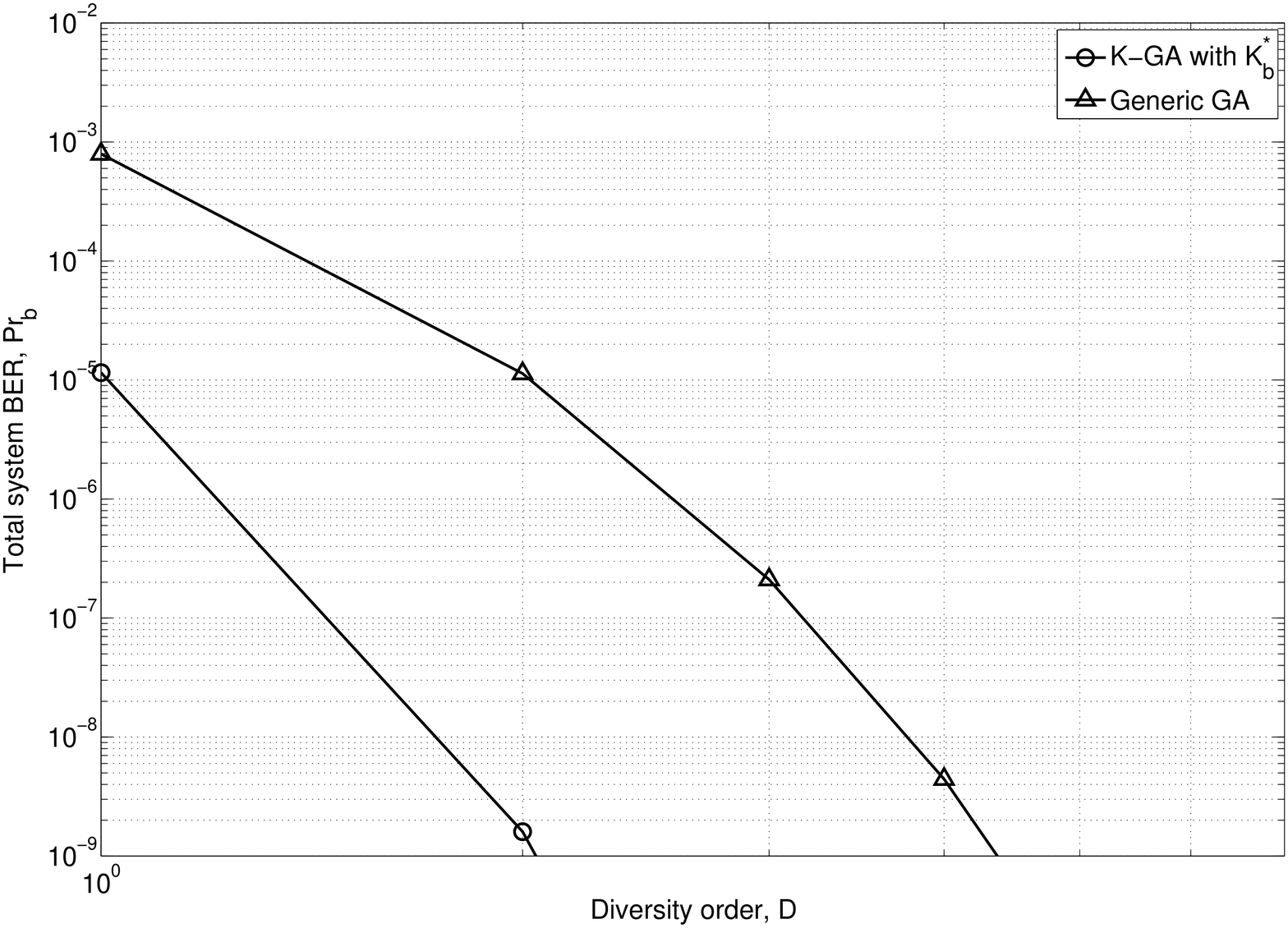}
\caption{Impacts of $D,$ i.e., $2D$ is the DOF resulting from
multiple antennas, on ${\rm Pr}_b^{K-GA}$ using $K^*_b$ when
${\bar K}=5,$ $\Omega_N=5$~dB, and $M=4.$}
\label{fig:avg_Pbmimo_aga}
\end{figure}

For the multiple antenna case, Fig.~\ref{fig:avg_Pbmimo_aga} shows
${\rm Pr}_b^{K-GA}(\cdot)$ versus the diversity order $D$ for the
following parameters $\bar{K}=5$, $\Omega_N=5$~dB, $\alpha=1,$ and
$M=4$. It is also assumed that the energy is distributed according
to (\ref{eq:opt_prob1_minPb}) with ${\rm Pr}^{K-GA}_b (\cdot)$ as
derived in Section~\ref{sec:ext_mu}. It can be seen from this
figure that our optimal energy distribution gives a boost in the
system BER as compared to the generic GA scheduling scheme.

\section{Conclusions}
\label{sec:con} A new realistic energy model which describes the
distribution of the total finite users' energy between scheduling
and data transmission stages is developed for the energy limited
uplink MU wireless systems. MU scheduling algorithms which
maximize the MUDiv gain are derived for the aforementioned systems
to (i)~minimize the overall system BER for a fixed total energy of
all users in the system or (ii)~minimize the total energy of all
users for fixed BER requirements. It is shown that for a fixed
number of available users, an achievable MUDiv gain can be
improved by activating only a subset of users from the entire set
of users. Using asymptotic analysis, it is shown that our approach
benefits from MUDiv gains higher than that achieved by the generic
GA algorithm, which is the optimal scheduling method for energy
unlimited systems. In particular, when minimizing the system BER,
it is found that the achieved MUDiv gain is determined by
$\left((\bar{K}-K)\alpha^{-1}+1 \right)\cdot{\rm SNR}$ when $K$ is
large. Moreover, in the case of high SNR, the MUDiv power gain
$\left((\bar{K}-K)\alpha^{-1}+1 \right)^{-K}$ can be archived
while obtaining the diversity order $K$. Simulation results
validate our theoretical observations and show that the proposed
$K$-GA algorithm based on optimizing the number of active users
provides significant energy gains for energy limited MU wireless
systems over the generic GA algorithm.

\section*{Appendix: Derivations of (\ref{eq:sol_optKb}) and
(\ref{eq:sol_opt_prob1})} \label{sec:appenx3} Using
(\ref{eq:apPb_ga}), the first derivative of ${\rm
Pr}_{b,a}^{K-GA} (K)$ in the optimization problem
(\ref{eq:opt_prob1-2_minPb}) can be expressed as
\begin{equation}
\frac{\partial}{\partial K}{\rm Pr}_{b,a}^{K-GA}\left(K \right) =
b {\rm B}(K,1+f(K)) + b K \frac{\partial}{\partial K}{\rm
B}\left(K,1+f(K)\right). \label{eq:diff_Pb}
\end{equation}
In turn, the first derivative of ${\rm B}\left(K,1+f(K)\right)$
with respect to $K$ in (\ref{eq:diff_Pb}) can be written as
\begin{equation}
\frac{\partial}{\partial{K}} {\rm
B}\left(K,1+f(K)\right)=\frac{\partial}{\partial{K}}\int_0^1
t^{K-1} (1-t)^{f(K)}{\rm d}t = \int_0^1
\frac{\partial}{\partial{K}} \hspace{0.1cm} t^{K-1} (1-t)^{f(K)}
{\rm d}t. \label{eq:diff_beta_0}
\end{equation}
or equivalently as
\begin{equation}
\frac{\partial}{\partial K} {\rm B}(K,1+f(K)) = \int_0^1
\left(1-t\right)^{f(K)} t^{K-1} \hspace{0.1cm} {\rm ln}t
\hspace{0.1cm} {\mathrm d}t + f'(K)\int_0^1 \left(1-t\right)^{K-1}
t^{f(K)} \hspace{0.1cm} {\rm ln}\hspace{0.05cm} t \hspace{0.1cm}
{\mathrm d}t \label{eq:diff_beta}
\end{equation}
where $f'(\cdot)$ denotes the first derivative of $f(\cdot)$ with
respect to $K$ and ${\rm ln}(\cdot)$ stands for the natural
logarithm\footnote{Note that a logarithm with any basis can
replace the natural logarithm.}.

Using the relationship \cite[(4.253.1)]{table_integ}
\begin{equation}
\int_0^1 x^{u-1} (1-x^r)^{v-1} {\rm ln}\hspace{0.05cm}x
\hspace{0.1cm} {\rm d}x = {\rm B}\left(u/r,v \right)
\big\{\psi(u/r)-\psi(u/r+v) \big\} / r^2 \label{eq:lem3}
\end{equation}
where $\psi(z)=\frac{\partial}{\partial z}{\rm ln}\Gamma(z)$
denotes the digamma function for $z>0$, the first derivative of
the beta function in (\ref{eq:diff_beta}) can be written as
\begin{equation}
\begin{split}
\frac{\partial}{\partial K} {\rm B}(K,1+f(K)) &= {\rm B}(K,1+f(K)) \\
\quad &\times \big\{
\psi(K)-(1+f'(K))\psi(K+1+f(K))+f'(K)\psi(1+f(K)) \big\}.
\end{split}
\label{eq:diff_beta2}
\end{equation}
Inserting (\ref{eq:diff_beta2}) into (\ref{eq:diff_Pb}), we find
that the solution of (\ref{eq:opt_prob1-2_minPb}) should satisfy
the following equation
\begin{equation}
{\rm B}(K,1+f(K)) \Big[ 1 + K \big\{
\psi(K)-(1+f'(K))\psi(K+1+f(K))+f'(K)\psi(1+f(K)) \big\} \Big]=0.
\label{eq:diff_Pb_2}
\end{equation}
Since in our system model $K\geq1$ and $f(K)\geq0$, it follows
from (\ref{eq:diff_Pb_2}) that ${\rm B}(K,1+f(K))\geq 0$.
Therefore, the equality ${\rm B}(K,1+f(K))= 0$ holds if and only
if $K$ goes to infinity. However, for $K\rightarrow\infty$ the
assumption of the limited total system user energy is violated,
and therefore, ${\rm B}(K,1+f(K))$ in (\ref{eq:diff_Pb_2}) must
always be positive. Thus, the problem of finding the solution of
(\ref{eq:opt_prob1-2_minPb}) boils down to the problem of finding
the number of users which satisfies the following equation
\begin{equation}
\psi(K)-\psi(K+1+f(K)) + f'(K) \big\{ \psi(1+f(K)) -
\psi(K+1+f(K)) \big\}+1/K = 0. \label{eq:opt_prob1-3_minPb}
\end{equation}

Using the following expression \cite[(8.365.3)]{table_integ}
\begin{equation}
\psi(x+n)=\psi(x)+\sum_{l=0}^{n-1}(x+l)^{-1}
\end{equation}
the differences between the digamma functions in
(\ref{eq:opt_prob1-3_minPb}) can be represented alternatively as
\begin{align}
\psi(K+1+f(K))-\psi(1+f(K)) &= \sum_{l=0}^{K-1}\left( 1+f(K)+l
\right)^{-1}.
\label{eq:diff_digamma1}\\
\psi(K+1+f(K))-\psi(K) &= \sum_{l=0}^{f(K)} \left( K+l
\right)^{-1}. \label{eq:diff_digamma2}
\end{align}
Finally, inserting (\ref{eq:diff_digamma1}) and
(\ref{eq:diff_digamma2}) into (\ref{eq:opt_prob1-3_minPb}), the
left hand side of (\ref{eq:opt_prob1-3_minPb}) can be rewritten as
\begin{equation}
\eta(K) = \frac{1}{K} -
\sum_{l=0}^{f(K)}\frac{1}{K+l} -
f'(K)\sum_{l=0}^{K-1}\frac{1}{1+f(K)+l}.
\end{equation}
Therefore, for given $\Omega$ and $E_T^{K-GA}$, the optimization
problem (\ref{eq:opt_prob1_minPb}) can be rewritten as
\begin{equation}
K^*=\arg\min_{K\in[K_n,K_1] }   \big|   \eta(K) \big|
\hspace{0.2cm} \text{subject to} \hspace{0.2cm} E_T^{K-GA} = E .
\end{equation}
This completes the derivation.

\newpage

\end{document}